\documentclass[prb,twocolumn,amsmath,amssymb,letterpaper,fixfloats]{revtex4}

\usepackage[mathbf,mathcal]{euler} 

\usepackage{graphicx}
\graphicspath{{FIGS//}}

\begin{document}

\title{Calculating potentials of mean force and diffusion coefficients from
  nonequilibirum processes without Jarzynski's equality}

\author{Ioan Kosztin}
\thanks{Corresponding author}
\email{KosztinI@missouri.edu}
\affiliation{Department of Physics \& Astronomy, University of Missouri,
  Columbia, MO 65211}

\author{Bogdan Barz}
\affiliation{Department of Physics \& Astronomy, University of Missouri,
  Columbia, MO 65211}

\author{Lorant Janosi}
\affiliation{Department of Physics \& Astronomy, University of Missouri,
  Columbia, MO 65211}

\date{October 27, 2005}

\begin{abstract}

  In general, the direct application of the Jarzynski equality (JE) to reconstruct
  \emph{potentials of mean force} (PMFs) from a small number of nonequilibrium
  unidirectional \emph{steered molecular dynamics} (SMD) paths is hindered by the
  lack of sampling of extremely rare paths with negative dissipative work. Such
  trajectories, that transiently violate the \emph{second law}, are crucial for
  the validity of JE.
  As a solution to this daunting problem, we propose a simple and efficient
  method, referred to as the \emph{FR method}, for calculating simultaneously both
  the PMF $U(z)$ and the corresponding diffusion coefficient $D(z)$ along a
  reaction coordinate $z$ for a classical many particle system by employing a
  small number of fast SMD pullings in both forward (F) and time reverse (R)
  directions, without invoking JE.
  By employing Crook's \emph{transient fluctuation theorem} (that is more general
  than JE) and the stiff spring approximation, we show that: (i) the mean
  dissipative work $\overline{W}_d$ in the F and R pullings are equal, (ii) both
  $U(z)$ and $\overline{W}_d$ can be expressed in terms of the easily calculable
  mean work of the F and R processes, and (iii) $D(z)$ can be expressed in terms
  of the slope of $\overline{W}_d$.
  To test its viability, the FR method is applied to determine $U(z)$ and $D(z)$
  of single-file water molecules in single-walled carbon nanotubes (SWNTs). The
  obtained $U(z)$ is found to be in very good agreement with the results from
  other PMF calculation methods, e.g., umbrella sampling.
  Finally, $U(z)$ and $D(z)$ are used as input in a stochastic model, based on the
  Fokker-Planck equation, for describing water transport through SWNTs on a
  mesoscopic time scale that in general is inaccessible to MD simulations.

\end{abstract}


\maketitle

\section{Introduction}
\label{sec:intro}

The study of the structure-function relationship of large biomolecules often
requires to follow their dynamics on a meso- or even macro-scopic time scale while
retaining its atomic scale spatial resolution.  A typical example is molecular and
ion transport through channel proteins\cite{roux02-182}. While structural details
of the inner lining of the channel in particular, and that of the
protein-lipid-solvent environment in general, are needed at atomic resolution in
order to determine the forces that guide the diffusion of the particles across the
channel, the duration of the permeation process may exceed by several orders of
magnitude the time scale of several tens of nanoseconds currently attainable by
all atom molecular dynamics (MD) simulations\cite{CBB01}.
In this case a simplified alternative approach is to model the transported
molecule in the channel as an overdamped Brownian particle that diffuses along the
axis of the channel in the presence of an effective \emph{potential of mean force}
(PMF) that describes its interaction with the rest of the atoms in the
system\cite{zwanzig2001}.
A PMF is the Landau free energy profile along a \emph{reaction coordinate} (RC),
or \emph{order parameter}\cite{LEAC2001}, and it can be determined from the equilibrium
statistical distribution function of the system by systematically integrating out
all degrees of freedom except the RC\cite{frenkel2002}.
In principle, both the effective diffusion coefficient and the PMF, quantities
that enter the Langevin equation of motion (or, equivalently, the corresponding
Fokker-Planck equation \cite{risken96}) which determines the dynamics of the
transported molecule, can be determined from MD simulations.
In practice, however, the calculation of free energy differences and PMFs are
rather difficult and computationally expensive\cite{frenkel2002,simonson02-430}. 

Since even the longest equilibrium MD (EMD) trajectories can sample only a small
region of the RC domain of interest, the one situated in the vicinity of the
corresponding PMF minimum, simple EMD simulations are not suitable for PMF
calculations.
The traditional method for calculating PMFs by means of biased EMD simulations is
\emph{umbrella sampling} (US)\cite{frenkel2002,roux95-275,torrie77-187}. 
However, US may become inefficient and computationally unaffordable when the
number of required sampling windows becomes too large. This may happen when the
amplitude of the equilibrium fluctuations of the RC is very small compared to the
size of the RC interval in which the PMF is sought.  

In such cases the RC can be sampled efficiently by employing \emph{steered
  molecular dynamics}\cite{isralewitz01-13} (SMD) in which the system is guided
(or steered), according to a predefined protocol, along the RC by using, e.g., a
\emph{harmonic guiding potential} (HGP).
By choosing a sufficiently large value for the elastic constant of the HGP, i.e.,
within the \emph{stiff-spring} approximation\cite{jensen02-6731,park04-5946}
(SSA), the distance between the target and actual value of the RC at a given time
can be kept below a desired value.
In general, for a large system ($\sim 10^5$~atoms) computationally one can afford
only a limited number (typically $\lesssim 10$) of such \emph{nonequilibrium} SMD
pullings, and the real challenge is to find a way to reconstruct the PMF (at least
semiquantitatively) along the RC using this limited amount of data.
In principle, the equilibrium PMF can be reconstructed from the celebrated
\emph{Jarzynski equality} (JE) that relates the equilibrium free energy difference
$\Delta{F}$ between two states to the average of the external work $W$ done along
all nonequilibrium paths that connect those states and are subject to the
preestablished RC variation protocol\cite{jarzynski97-2690,jarzynski97-5018}.
In terms of the \emph{dissipative work} $W_d=W-\Delta{F}$, JE can be written as
$\left\langle \exp(-\beta W_d) \right\rangle = 1$, where $\beta=1/k_BT$, $k_B$ is
the Boltzmann constant and $T$ is the temperature of the heat bath (environment).
We note that if the SMD pulling occurs infinitely slowly then the system is in
equilibrium at all times and $W_d=0$ (reversible paths). Thus JE is trivially
satisfied and $W\equiv W_{rev}=\Delta F$ is the \emph{reversible work}.
In general SMD pullings are nonequilibrium with $W_d>0$ along most of the
trajectories. 
However, the validity of JE depends crucially on a small fraction of trajectories
with $W_d<0$, that transiently violate the \emph{second law}. Since such
trajectories (whose number decreases exponentially with $\overline{W}_d/k_BT$) are very
unlikely to occur among a few fast SMD pullings, it is clear that the sought PMF
cannot be determined by the direct application of JE, except when the pulling
paths are close to equilibrium (i.e., when $\overline{W}_d\lesssim k_BT$).
Under near-equilibrium conditions, the validity of JE has been confirmed in an RNA
stretching experiment\cite{liphardt02-1832}. Also, JE has been
successfully applied for free energy calculations in computer simulation of small
and/or simplified model systems.  However, in spite of a large number of papers
dedicated to the applications of
JE\cite{bustamante05-43,gore03-12564,ritort02-13544,hendrix01-5974,%
  hummer01-7330,hummer01-3658,hummer05-504,atilgan04-10392,sun03-5769,%
zuckerman02-180602,zuckerman02-445} and
other \emph{fluctuation theorems}\cite{evans02-1529,wang02-050601}, there are
surprisingly few studies which use SMD simulations combined with the JE to
calculate PMFs for large biomolecules\cite{jensen02-6731,vidossich04-924}.

The purpose of this paper is to propose a simple and efficient method, referred to
as the \emph{FR method}, for calculating simultaneously both the PMF $U(z)$ and
the corresponding diffusion coefficient $D(z)$ along a reaction coordinate $z$ for
a classical many particle system by employing a small number of fast
nonequilibrium SMD pullings in both forward (F) and time reverse (R) directions,
without invoking JE. In fact, as already mentioned, for such limited number of
processes JE fails to hold.
The essence of the FR method, detailed in Sec.~\ref{sec:theory}, can be summarized
as follows: 
Several fast F and R SMD pullings are carried out within the SSA. The latter
guaranties that (i) the RC follows closely its target value determined by the
pulling protocol, (ii) the change in PMF ($\Delta{U}$) is well approximated by the
corresponding change in the free energy ($\Delta{F}$) of the system biased by the
HGP, and (iii) the work distribution function $P_{F/R}(W)$ along F/R paths is
Gaussian.
A few F and R SMD trajectories are sufficient to sample $P_{F/R}(W)$ about its
maximum (see Fig.~\ref{fig:p-w}) and, therefore, determine approximately the mean
F/R work $\overline{W}_{F/R}$. However, the same data is insufficient for even a
rough estimate of the variance $\sigma^2_{W}=\overline{W^2} - \overline{W}^2$,
i.e., of the actual width of $P_{F/R}(W)$.
From Crooks' \emph{transient fluctuation theorem}\cite{crooks00-2361} (TFT) [see
Eq.~\eqref{eq:tft}], which is more general than JE, follows that if $P_F(W)$ is
Gaussian then $P_R(W)$ is also a Gaussian with the same variance (width)
$\sigma^2_W = 2k_BT \, \overline{W}_d$, and peak position
$\overline{W}_R=\overline{W}_F-2\Delta{F}$.
Thus, (i) the PMF is given by $\Delta{U}=\Delta{F}=(\overline{W}_F -
\overline{W}_R)/2$, and (ii) the mean dissipative work is the same for both F and
R paths, given by $\overline{W}_d = (\overline{W}_F + \overline{W}_R)/2$.
From $\overline{W}_d$ the position dependent diffusion coefficient is $D=k_{B}T
v/(d\overline{W}_d/dz$), where $v$ is the pulling speed.

Thus, the reason why previous studies failed to reconstruct the PMF from
\emph{unidirectional} SMD pullings far from nonequilibrium by using JE is because
such approach requires the complete sampling of the corresponding work
distribution function that is simply impossible to obtain from a limited number of
pullings. (In fact, the sampling of $P(W)$ has to be so complete that it must
include paths with $W_d<0$ as discussed above.) While the mean work can be easily
estimated, breaking this up into the PMF and the mean dissipative work (i.e., the
heat exchanged with the environment) requires either the knowledge of the precise
width (variance) of the F work distribution function (e.g., when the F SMD paths
are close to equilibrium and $\overline{W}_d$ is small) or additional information
that may come from a set of R SMD pullings, as outlined above.
The solution to this problem offered by our FR method is surprisingly simple,
however, its validity depends crucially on Crooks' TFT (from which JE can be
derived) and the Gaussian nature of $P_{F/R}(W)$ guaranteed by SSA. In particular,
the conclusion that the mean dissipative work is the same for both F and R SMD
paths is highly non trivial.

The remaining of the paper is organized as follows. In Sec.~\ref{sec:theory} we
describe in detail the theoretical basis of our proposed FR method.In order to
test its efficiency and viability, in Sec.~\ref{sec:res} we apply the FR method to
calculate the PMF and the position dependent diffusion coefficient of water
molecules moving across densely packed single walled carbon nanotubes (SWNT) that
connect two water reservoirs.  The obtained PMF is compared with the ones obtained
from EMD and US simulations.  Finally, conclusions are drawn in
Sec.~\ref{sec:conc}.

\section{Theory}
\label{sec:theory}
We consider a classical many particle system (e.g., a channel protein in a fully
solvated lipid bilayer) described by the Hamiltonian $H_0(\Gamma)$, where
$\Gamma\equiv\{\mathbf{r},\mathbf{p}\}$ represents the coordinates and momenta of
all the atoms in the system. The dynamics of the system may be either
deterministic or stochastic, but we assume that the conditions for which JE and
TFT hold are met, i.e., the dynamics are Markovian and preserve the equilibrium
ensemble, and the energy of the system is finite\cite{crooks00-2361}. These
conditions are met in MD simulations in both NVT and NpT
ensembles\cite{park04-5946}.

\subsection{Reaction coordinate and PMF}
\label{sec:rc-pmf}
In general, any PMF calculation starts with the identification of a properly
chosen RC whose change in time describes the evolution of the state of the
system\cite{frenkel2002}.  For example, in describing the progression of a
transported molecule in a nanopore (e.g., channel protein or SWNT) a proper RC is
the projection of the COM of the molecule (or of a part of the molecule) on either
the permeation direction across the membrane, hereafter denoted by $z$, or on the
axis of the pore. If the channel is relatively straight then the difference
between the two RC choices can be neglected. For simplicity, here we assume that
this is always the case.

\begin{figure}[Ht]
  \centering
  \includegraphics[width=3.2in]{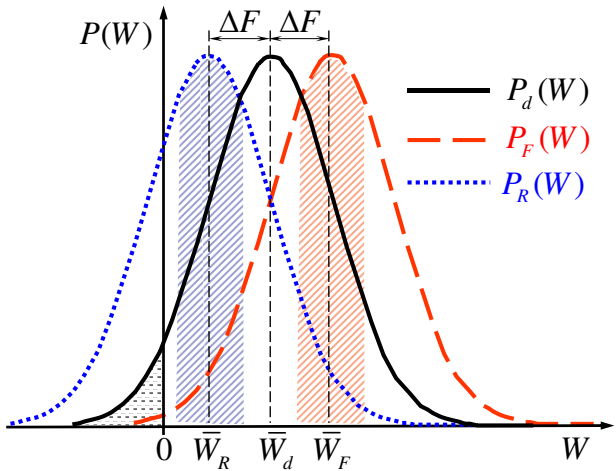}  
  \caption{(Color online) Gaussian forward (long-dashed), reverse (dotted) and
    dissipative (solid) work distribution functions within the \emph{stiff spring}
    approximation. The shaded region in $P_{F/R}(W)$ is the one sampled in F/R SMD
    pullings. The tail region of $P_d(W)$ corresponding to negative dissipative
    work is also highlighted.}
  \label{fig:p-w}
\end{figure}

By definition, the PMF $U(z)$ is determined from the equilibrium distribution
function of the system by integrating out all degrees of freedom except the
reaction coordinate $z$, i.e., \cite{frenkel2002}
\begin{equation}
  \label{eq:pmf1}
  e^{-\beta U(z)}\equiv p_0(z)=\int d\Gamma \frac{e^{-\beta H_0(\Gamma)}}{Z_0}
  \delta[z-\tilde{z}(\Gamma)] \;,
\end{equation}
where $p_0(z)$ is the equilibrium distribution function of the reaction
coordinate, $Z_0$ is the partition function and $\delta(z)$ is the Dirac-delta
function whose filtering property guarantees that the integrand in
Eq.\eqref{eq:pmf1} is nonzero only when the RC has the desired value, i.e, when
$\tilde{z}(\Gamma)=z$. Hereafter we use the convention that $z$ [or $z(t)$]
represents the target value of the RC while $\tilde{z}\equiv \tilde{z}(\Gamma)$
represents the actual value of the RC.
Also, unless otherwise stated, throughout this paper the energy is measured in
units of $k_B T$, e.g., in Eq.~\eqref{eq:pmf1} one needs to set $\beta=1$.

In principle, the equilibrium distribution function $p_0(z)$ can be easily
computed from EMD simulations, since it is proportional to the logarithm of the
binned histogram of the RC sampled along the MD trajectory. Thus, the PMF is
readily given by
\begin{equation}
  \label{eq:pmf0}
U(z)=-\log \left[ p_0(z) \right]\;.  
\end{equation}
In terms of the $U(z)$ the equilibrium average of any function $f(\tilde{z})$
of the RC can be calculated as

\begin{eqnarray}
  \label{eq:pmf-mean}
  \left\langle f(\tilde{z}) \right\rangle_0 &=& \int d\Gamma
  \frac{e^{-H_0(\Gamma)}}{Z_0} f(\tilde{z}) \int dz\, \delta[z-\tilde{z}(\Gamma)]\\
  &=& \int dz\, e^{-U(z)} f(z)  = \int dz\, p_0(z) f(z) \;. \nonumber
\end{eqnarray}

In practice, however, even the longest EMD trajectories sample only a restricted
region of the reaction coordinate domain of interest (i.e., within the vicinity of
the PMF minimum) and the direct application of Eq.~\eqref{eq:pmf1} is impractical.
\subsection{Harmonic guiding potential}
\label{sec:hgp}
In order to properly sample energetically more difficult to reach regions, one
needs to \emph{guide} or \emph{steer} the system towards those regions by
employing, e.g., a harmonic guiding potential (HGP)
\begin{equation}
  \label{eq:hgp}
  V_z(\tilde{z})\equiv V(\tilde{z}(\Gamma)|z) = \frac{k}{2}[\tilde{z}(\Gamma)-z]^2 \;,
\end{equation}
where $k\equiv k_z$ is the stiffness (elastic constant) of the HGP. With this extra
potential energy, the Hamiltonian of the new biased system becomes $H_z = H_0 +
V_z(\tilde{z})$.  As a result, atom ``$j$'' in the selection that define the
RC will experience an additional force
\begin{equation}
  \label{eq:hgp2}
\mathbf{F}_j =
-\frac{\partial{V_z}}{\partial{\mathbf{r}_j}}=-k  [\tilde{z}(\Gamma)-z]
\frac{\partial{\tilde{z}(\Gamma)}}{\partial{\mathbf{r}_j}}\;. 
\end{equation}
Thus, the HGP (\ref{eq:hgp}) will force the system to evolve in the configuration
space in such a way that at all times $\tilde{z}$ stays confined in the vicinity
of $z$.

The free energy difference $\delta F_z=F_z-F_0$ between the equilibrium states of
the systems described by the Hamiltonians $H_z$ and $H_0$ can then be written as a
Gaussian convolution of $exp[-U(z)]$. Indeed,

\begin{eqnarray}
  \label{eq:hgp3}
  e^{-\delta F_z} &=& \int d\Gamma \frac{e^{-\beta H_0(\Gamma)}}{Z_0}
  e^{-V_z(\tilde{z}(\Gamma))} = \left\langle e^{V_z(\tilde{z})} \right\rangle_0 \\
  &=& \int dz' e^{-U(z')} e^{-V_z(z')} = \int dz' e^{-U(z')}
  e^{-\frac{k}{2}(z-z')^2}\;. \nonumber
\end{eqnarray}
\subsection{Stiff-spring approximation}
\label{sec:ssa}
The sought PMF, $U(z)$, can be obtained from Eq.~\eqref{eq:hgp3} by Gaussian
deconvolution of the free energy factor $\exp(-\delta F_z)$. However, it is more
convenient to resort to the large $k$ or \emph{stiff-spring
  approximation}\cite{park04-5946,park03-3559,jensen02-6731} (SSA).  Assuming that
we seek to determine $U(z)$ with a spatial resolution $\delta{z}$, by choosing the
spring constant such that $k\gg 2/(\delta{z})^2$ one can easily see that in
Eq.~\eqref{eq:hgp3} the main contribution to the last integral comes from the
region $|z-z'|\ll \delta{z}$, and therefore one can write
\begin{equation}
  \label{eq:ssa1}
  e^{-\delta F_z} \approx e^{-U(z)} \int dz' e^{-\frac{k}{2}(z-z')^2} =
  \sqrt{\frac{2\pi}{k}} e^{-U(z)} \;.
\end{equation}
Now, taking the logarithm of both sides in Eq.~\eqref{eq:ssa1}, one obtains $\delta
F_z = F_z-F_0=U(z)+\text{const}$ and, therefore,
\begin{equation}
  \label{eq:ssa2}
  \Delta{U}=U(z)-U(z_0) \approx \Delta{F}=F_z-F_{z_0}\;.
\end{equation}
Thus, within the SSA the PMF of the unbiased system is well approximated by the free
energy difference of the system biased by the HGP. 
Note that in SMD simulations, to make sure that the distance between the target
$z(t)$ and actual $\tilde{z}$ values of the RC on average stays smaller than the
desired $\delta{z}$, one needs to chose the spring constant according to 
\begin{equation}
  \label{eq:ssa3}
  k \gtrsim \text{max}\left\{ \frac{2\alpha}{(\delta{z})^2},
    \frac{2U_{max}}{(\delta{z})^2} \right\}\;, 
\end{equation}
where $U_{max}$ is the highest PMF barrier one wants to explore, and $\alpha\gg
1$.

\subsection{PMF from umbrella sampling  and WHAM}
\label{sec:us-wham}
In umbrella sampling\cite{roux95-275,kumar92-1011,torrie77-187,frenkel2002}, the
range of RC values of interest $(z_{min},z_{max})$ is divided into $N_w$
\emph{sampling windows} centered about conveniently chosen values $z_i$,
$i=1,\ldots,N_w$. Next, the reaction coordinate is sampled in each window
separately by preparing identical replicas of the system and applying the harmonic
guiding potential $V_{z_i}(\tilde{z})$. As a result, the biased distribution
functions can be readily obtained by direct sampling of the reaction coordinate
for the biased system, i.e, $p_i(z) = \left( Z_0/Z_i \right) e^{- V_i(z)} p_0(z)$,
where, for brevity, the index $z_i$ has been replaced by $i$. By inverting this
equation, the equilibrium distribution in each window can be expressed in terms of
the biased distribution of the reaction coordinate.
The standard method for efficiently stitching together the biased $p_i(z)$'s in order to obtain
the equilibrium $p_0(z)$, and therefore the sought PMF, is the so called
\emph{weighted histogram analysis method} or WHAM \cite{kumar92-1011}, according
to which 
\begin{subequations}
  \label{eq:wham}
\begin{eqnarray}
  \label{eq:wham1}
  p_0(z) &=& \frac{\sum_{i=1}^{N_w} \mathcal{N}_i p_i(z)}{\sum_{i=1}^{N_w}
    \mathcal{N}_i e^{-V_i(z)}/\langle e^{-V_i} \rangle}
  \;,\\
  \label{eq:wham2}
 \langle e^{-V_i} \rangle &=& \int\!\! dz\, p_0(z) e^{-V_i(z)} \;,
\end{eqnarray}
\end{subequations}
with $\mathcal{N}_i$ the number of data points used to construct $p_i(z)$.  The
above non-linear coupled WHAM equations, that need to be solved iteratively,
minimize the errors in determining $p_0(z)$. 
When applicable, US combined with WHAM is perhaps the best choice for calculating
PMFs. In practice, however, one often encounters situations in which the minimum
number of US windows required to properly cover the range of RC values of interest
is excessively large and the application of the method may become computationally
unattainable. Molecular transport in channel proteins is a good example.

\subsection{SMD, Transient Fluctuation Theorem  and the Jarzynski equality}
\label{sec:smd-tft-je}
In SMD simulations\cite{isralewitz01-13}, where initially the system is in an
equilibrium state characterized by $z(0)$, the target value of the RC $z(t)$ (also
referred to as \emph{control parameter}) is varied in time according to a
prescribed protocol.  For example, in constant velocity SMD (cv-SMD)
$z(t)=z(0)+vt$, $0\le t\le\tau$, where $v$ is the constant pulling speed equal to
the ratio of the total pulling distance to the desired simulation time. We refer
to the SMD pulling paths of the system when $t$ increases from $0$ to $\tau$ as
forward (F) paths. The time reverse (R) pulling paths are obtained by starting the
system from an equilibrium state corresponding to $z(\tau)$ and reversing the sign
of $t$ in $z(t)$ for F paths. In our cv-SMD example, this amounts to setting
$z_R(t)=z_F(\tau-t)=z(\tau)-vt$, $0\le t\le\tau$.
The choice of a sufficiently large spring constant (see Sec~\ref{sec:ssa}) in the
now time dependent HGP [Eq.~\eqref{eq:hgp}] guarantees that the instantaneous RC,
$\tilde{z}(t)$, follows closely the target value $z(t)$ during the pulling
process. Thus, cv-SMD is a fast sampling method of the RC by driving the system
out of equilibrium. The faster the pulling the more significant is the deviation
from equilibrium. 
The work done during a cv-SMD simulation is given by
\begin{equation}
  \label{eq:work}
W_t\equiv W_z=\int_{z_0}^{z(t)} dz
\left[ \partial{V_z(\tilde{z})}/\partial{z}\right] = k\int_{z_0}^{z(t)}
dz (z-\tilde{z})\;. 
\end{equation}

Crooks has shown that under rather general conditions, listed at the beginning of
this section, the following nonequilibrium fluctuation theorem holds\cite{crooks00-2361}
\begin{subequations}
\label{eq:crooks}
\begin{equation}
  \label{eq:crooks-F}
  \left\langle f(W)e^{-W_{dF}} \right\rangle_{F} = \left\langle f(-W) \right\rangle_{R}\;,
\end{equation}
or
\begin{equation}
  \label{eq:crooks-R}
  \left\langle f(W) \right\rangle_{F} = \left\langle f(-W) e^{-W_{dR}}\right\rangle_{R}\;.
\end{equation}
\end{subequations}
Here $f(W)$ is an arbitrary function of the work $W$, and
\begin{equation}
  \label{eq:smd-tft-je1}
  \left\langle \ldots \right\rangle_{F/R} = \int dW\, P_{F/R}(W) \ldots \;,
\end{equation}
represents the average over forward/reverse paths or, equivalently, the average
with respect to the forward/reverse work distribution functions $P_{F/R}(W)$.
The dissipative work in a F/R process is given by
  \begin{equation}
    \label{eq:wdF}
    W_{d\,F/R}=W_{F/R}\mp\Delta{F}\;,
  \end{equation}
with $\Delta{F}=F_{z(\tau)}-F_{z(0)}$.
The JE follows immediately from Eqs.~\eqref{eq:crooks} by setting
$f(W)=1$, and it can be written in any of the following forms
\begin{subequations}
\label{eq:JEs}  
\begin{equation}
  \label{eq:JE1}
  \left\langle \exp(-W_{dF}) \right\rangle_F = \left\langle \exp(-W_{dR}) \right\rangle_R = 1\;,
\end{equation}
\begin{equation}
  \label{eq:JE2}
  \left\langle \exp(-W) \right\rangle_F = e^{-\Delta{F}}\;,\;
   \left\langle \exp(-W) \right\rangle_R = e^{\Delta{F}}  \;.
\end{equation}
\end{subequations}
Another important equality, that connects the F and R work distribution functions,
can be derived from Eqs.~\eqref{eq:crooks} by setting $f(W')=\delta(W-W')$ and
carrying out the integral with respect to $W'$. The result is Crooks' transient
fluctuation theorem\cite{crooks00-2361} (TFT)
\begin{equation}
  \label{eq:tft}
  \frac{P_F(W)}{P_R(-W)}=e^{W_{dF}}\;.
\end{equation}
This equation is used to derive our new results in Sec.~\ref{sec:f-r-smd}.

\subsection{PMF from unidirectional SMD and the Jarzynski equality}
\label{sec:pmf-smd-je}
An increasingly popular alternative for calculating PMFs is based on the
application of the JE from repeated unidirectional nonequilibrium SMD simulations\cite{park04-5946,park03-3559,jensen02-6731,hummer05-504,hummer01-7330,hummer01-3658,vidossich04-924,cascella02-13027,amaro03-7599,amaro04-147}.
Within the SSA the sought PMF can be readily obtained from Eqs.~\eqref{eq:ssa2}
and ~\eqref{eq:JE2}
\begin{equation}
  \label{eq:pmf-smd-1}
  \Delta{U(z)}\approx\Delta{F}=-log \left\langle \exp(-W_z) \right\rangle_F\;.
\end{equation}
Here the index $F$ indicates that the average is taken over the ensemble of
forward pulling paths.
As already mentioned, the average of the exponential in Eq.~\eqref{eq:pmf-smd-1}
cannot be estimated reliably even for a reasonably large number of SMD pullings,
unless the pulling speed is sufficiently small so that the system is close to
equilibrium along the pulling paths. This is due to the fact that the overlap
between $\exp(-W)$ and the sampled part of $P_F(W)$ is in general exponentially
small. Nevertheless, there exist two approaches that in principle may give fairly
good estimates of Eq.~\eqref{eq:pmf-smd-1}, provided that the system is not too
far from equilibrium during pullings.
The first method is the cumulant
approximation\cite{park04-5946,park03-3559,jensen02-6731}, according to which
\begin{subequations}
\label{eq:pmf-smd-2}
\begin{eqnarray}
  \label{eq:pmf-smd-2a}
  \Delta{U(z)}&=&-\log \left\langle \exp(-W_z) \right\rangle \approx \overline{W}_z
   -\sigma^2_z /2\;,\\
  \sigma^2_z &=&  \overline{W_z^2} - \overline{W}_z^2\;,
\end{eqnarray}
\end{subequations}
where for simplicity we have dropped the index ``F'' and $\sigma^2_z$ is the
variance (2nd cumulant) of the work. It has been shown that within SSA the work
distribution function $P_F(W)$ is Gaussian, and therefore generally recognized
that in this case the cumulant approximation \eqref{eq:pmf-smd-2} in fact is
exact. However, as mentioned in Sec.~\ref{sec:intro}, the reason why in practice
Eq.~\eqref{eq:pmf-smd-2} is valid only close to equilibrium is because SMD pulling
paths can sample only a narrow region about the peak of the Gaussian $P_F(W)$.
This allows for a fairly accurate determination of the mean work $\left\langle W_z
\right\rangle$ but, in general seriously underestimates the variance $\sigma_z^2$.

The second method for evaluating the average in Eq.~\eqref{eq:pmf-smd-1} is a
weighted histogram approach suggested by Hummer and
Szabo\cite{hummer05-504,hummer01-3658}, and indirectly by
Crooks\cite{crooks00-2361}. The nonequilibrium fluctuation theorem due to Crooks
can also be written as
\begin{equation}
  \label{eq:crooks-neft2}
  \left\langle f[z(t)]\exp(-W_d) \right\rangle_F = \left\langle f[z_R(0)]
  \right\rangle_R = \left\langle f[z(t)] \right\rangle_{eq}\;
\end{equation}
where $z_R(t)$ represents the time evolution of the control parameter during
reverse pullings, $f[z]$ is an arbitrary function and the index ``eq''
means the equilibrium average corresponding to the biased system with Hamiltonian
$H_{z_0}$. By inserting $f[z]=\delta(z-\tilde{z})$ into
Eq.~\eqref{eq:crooks-neft2} one obtains 
\begin{eqnarray}
  \label{eq:pmf-smd-3}
   \left\langle \delta(z-\tilde{z}) e^{-W_{z'}} \right\rangle_F &=& \frac{Z_0}{Z_{z_0}}
   \left\langle \delta(z-\tilde{z}) e^{-V_{z'}(\tilde{z})} \right\rangle_0 \\
   \nonumber
  &=& \frac{e^{-V_{z'}(z)}}{\left\langle e^{-V_{z_0}} \right\rangle_0} e^{-U(z)}\;,
\end{eqnarray}
Since the equilibrium average $\left\langle \exp(-V_{z_0}) \right\rangle_0$
corresponding to the unbiased system contributes only an additive constant to the
PMF, from Eq.~\eqref{eq:pmf-smd-3} one obtains the following result
\begin{subequations}
\label{eq:pmf-smd-wh}
\begin{equation}
  \label{eq:pmf-smd-wh1}
  U(z) = -log \left\langle \delta(z-\tilde{z}) \exp(-\Delta W_{z'}) \right\rangle\;,
\end{equation}
where
\begin{eqnarray}
  \label{eq:pmf-smd-wh2}
  \Delta W_{z'} &=& W_{z'}-V_{z'}(\tilde{z})
  = k\int_0^t d\tau\, \dot{z'}(\tau)
  [z'(\tau)-\tilde{z}(\tau)] \nonumber \\
&&- \frac{k}{2}[z'(\tau)-\tilde{z}(\tau)]^2\;. 
\end{eqnarray}
\end{subequations}

Thus, $U(z)$ can be calculated from the work time series obtained in repeated
cv-SMD simulations by constructing a weighted histogram of the RC according to
Eqs.~\eqref{eq:pmf-smd-wh}.
This method resembles to the US and WHAM and is preferable to the cumulant
approximation method whenever we have a large number of pulling paths. 
However, in the case of large systems when only a limited number of trajectories
can be sampled this method is inapplicable because of insufficient data.

\subsection{PMF from forward and reverse SMD pullings with a stiff spring}
\label{sec:f-r-smd}
In this section we present our new method for calculating PMFs from few fast SMD
pullings along the RC in both F and R directions, hereafter referred to as the
\emph{FR method}. We assume that the pullings are done with a sufficiently stiff
spring such that the SSA holds (Sec.~\ref{sec:ssa}). In this case, the F work
distribution $P_F(W)$ is Gaussian, and according to Crooks' TFT \eqref{eq:tft} it
follows that the R work distribution $P_R(W)$ is also Gaussian. Thus one can
write
\begin{equation}
  \label{eq:fr-1}
  P_{F/R}(W)=\left( 2\pi\sigma_{F/R}^2 \right)^{\frac{1}{2}} \exp \left[
    -\frac{(W-\overline{W}_{F/R})^2}{2 \sigma_{F/R}^2}\right]
\end{equation}
where $\overline{W}_{F/R}$ and $\sigma_{F/R}^2$ are the mean work and variance
corresponding to the F and R pulling directions, respectively.
The mean dissipative work in the two distinct pulling directions is
\begin{equation}
  \label{eq:fr-dW}
  \overline{W}_{d\,F/R} = \int dW  (W\pm\Delta{F}) P_{F/R}(W) =
  \overline{W}_{F/R}\pm\Delta{F}\;. 
\end{equation}

Inserting \eqref{eq:fr-1} into  \eqref{eq:tft} and taking into account that
$W_{dF}=W-\Delta{F}$, after little algebra it follows that TFT can hold only if
\begin{subequations}
\label{eq:fr-2}
  \begin{equation}
    \label{eq:fr-2a}
    \sigma^2\equiv\sigma_F^{2}=\sigma_R^2 = \overline{W}_F + \overline{W}_R\;
  \end{equation}
and
\begin{equation}
  \label{eq:fr-2b}
  \Delta{F}=(\overline{W}_F-\overline{W}_R)/2\;.
\end{equation}
Finally, inserting Eq.~\eqref{eq:fr-2a} into \eqref{eq:fr-dW}, one finds that the
mean dissipative work is the same in both F and R pulling directions, i.e.,
\begin{equation}
  \label{eq:fr-2c}
  \overline{W}_d\equiv \overline{W}_{dF} = \overline{W}_{dR} =
  (\overline{W}_F+\overline{W}_R)/2 \;.
\end{equation}
\end{subequations}
Equations \eqref{eq:fr-2} are the key formulas of our FR method for calculating PMFs
from fast F and R SMD pullings. 
Assuming that a few ($\sim 10)$ such SMD pullings can sample reasonably well the
work about the peak position $\overline{W}_{F/R}$ of $P_{F/R}(W)$, as indicated by
the shaded regions in Fig.~\ref{fig:p-w}, then Eqs.~\eqref{eq:fr-2} yield
essentially with the same degree of accuracy both the desired PMF,
$\Delta{U}\approx\Delta{F}$, and the mean dissipative work, $\overline{W}_d$. This
feature makes the proposed method superior to the currently used approaches
described in the previous sections. In fact, these other methods can only
determine the mean total work $\overline{W}_F$ with some statistical correction
either through the cumulant approximation or a weighted histogram method.
Furthermore, since it is reasonable to assume that $\overline{W}_d$ is
proportional to the pulling speed $v$, one can readily determine the position
dependent friction coefficient $\gamma(z)$ from the slope of the mean dissipative
work $\gamma(z) =\left( d\overline{W}_d(z)/dz \right)/v$. Then, the corresponding
diffusion coefficient is given by the Einstein relation (in $k_BT$ energy units)
\begin{equation}
  \label{eq:fr-diff}
  D(z) = \gamma(z)^{-1}=v \left( d\overline{W}_d(z)/dz \right)^{-1} \;.
\end{equation}
Now that both $U(z)$ and $D(z)$ are determined, the equation of motion of the RC
on a meso (or macro) time scale is given by the Langevin equation corresponding to
an overdamped Brownian particle\cite{zwanzig2001}
\begin{subequations}
\label{eq:fr-L-FP}
  \begin{equation}
    \label{eq:fr-L}
    \gamma(z) \dot{z} = -dU(z)/dz + \xi(t)\;,
  \end{equation}
or equivalently, the corresponding Fokker-Planck equation for the probability
distribution function $p(z,t)$ of the RC 
  \begin{equation}
    \label{eq:fr-FP}
    \partial_t p(z,t) = -\partial_z j(z,t) = \partial_z D(z) \partial_z p(z,t) +
    \partial_z U'(z) p(z,t)\;,
  \end{equation}
\end{subequations}
where $\xi(t)$ is the Langevin force (modeled as a Gaussian white noise) and
$j(z,t)$ is the probability current density.

We emphasize again that far from equilibrium the variance $\sigma_W^2 \equiv
\sigma_z^2$ of the F/R work calculated from SMD pullings data
[cf.~\eqref{eq:pmf-smd-2}] is in general much smaller than the variance $\sigma^2$
of the actual work distribution function, and therefore it cannot be used to
estimate even approximately the mean dissipative work, unless an exponentially
large number of SMD trajectories are collected and used for this purpose.

Finally, we note that $P_F(W)$ and $P_R(W)$ are identical Gaussians centered about
$\overline{W}_F$ and $\overline{W}_R$, respectively. One can also define a
distribution function for the dissipative work through
$P_d(W)=P_F(W+\Delta{F})=P_R(W-\Delta{F})$, which is centered about
$\overline{W}_d$ (Fig.~\ref{fig:p-w}). This allows us to calculate the fraction of
the SMD trajectories that violates the second law, i.e., for which $W_d<0$; these
trajectories are crucial in establishing the validity of the JE. We have
\begin{eqnarray}
  \label{eq:2}
  \left\langle e^{-W_d} \right\rangle|_{W_d<0}  &=&\int_{-\infty}^{0} dW\, P_{d}(W)
  e^{-W}\\
 &=& \frac{1}{2} \text{erfc}\left(\overline{W}_d^{1/2}\right) \sim
  \frac{\exp(-\overline{W}_d)}{W_d^{1/2}}\;, \nonumber
\end{eqnarray}
which clearly indicates that for $\overline{W}_d > 1$ (i.e., $\overline{W}_d >
k_BT$ in SI units) the number of such trajectories is exponentially small, and
finding any of them in SMD simulations of large biomolecules is rather unlikely.

\subsection{Generalized acceptance ratio method}
\label{sec:arm}
The idea of combining results from both F and R simulations is not new, dating
back to the original Bennett's \emph{acceptance ratio method}\cite{bennett76-245}. 
However, in previous such
studies\cite{rodinger05-164,lu03-2977,lu04-173,lu04-05772,shirts03-140601,shirts05-144107} the
focus was mainly on determining the free energy difference between two states and
to estimate the corresponding error, unlike in our FR method in which the PMF, the
mean dissipative work and the corresponding diffusion coefficient are determined
simultaneously from specially designed F and R pullings with Gaussian distributed
work.
For example, starting from the nonequilibrium fluctuation theorem
\eqref{eq:crooks} and following the general philosophy of the Bennett acceptance
ratio method, Crooks has shown\cite{crooks00-2361} that the best estimate (i.e.,
with smallest error) of the free energy [see Eqs.~\eqref{eq:crooks}]
\begin{equation}
  \label{eq:ar-1}
  e^{-\Delta{F}}=\left\langle f(W) \right\rangle_F / \left\langle f(-W)e^{-W}
    \right\rangle_R
\end{equation}
is obtained by choosing the $f(W)=1/[1+n_F/n_R \exp(W-\Delta{F})]$, where
$n_{F/R}$ represent the number of F/R paths sampled. Essentially the same result
was derived by Pande and collaborators\cite{shirts03-140601} by applying the
\emph{maximum likelihood estimator} (MLE) method to Crooks' TFT \eqref{eq:tft}.
Thus, the best estimate of the free energy difference $\Delta{F}$ between two
equilibrium states corresponding to the RCs $z(0)$ and $z(t)$ is given by the
solution of the following transcendental equation
\begin{eqnarray}
  \label{eq:ar-2}
    &&\sum_{i=1}^{n_F}\frac{1}{1+n_F/n_R \exp(W_{F i}-\Delta{F})}\\
    && -
    \sum_{i=1}^{n_R}\frac{1}{1+n_R/n_F \exp(-W_{R i}-\Delta{F})} = 0 \;.\nonumber
\end{eqnarray}
In order to calculate the PMF $U(z)$ along the RC, $z$, by using the above MLE
method, first, one needs to divide the domain of interest $\{z_{min},z_{max}\}$
into $N$ intervals determined by the division points $z_i$, $i=0,\ldots,N$. Then
the system needs to be steered into these point via SMD, and in each of them it
needs to be equilibrated. Then, depending on the available computational
resources, a well defined number of F and R cv-SMD pullings should be carried out
between adjacent division point, each time starting from a different equilibrium
configuration.  Finally, solving Eq.~\eqref{eq:ar-2} within the SSA, one
determines the change $\Delta{U}_i=U_i-U_{i-1}$ along each segment
$(z_{i-1},z_i)$.
Although, strictly speaking, the above methods that combine F and R SMD pullings can
determine the free energy difference between initially equilibrated states, in
practice we find that in many cases Eqs~\eqref{eq:fr-2b}-\eqref{eq:fr-2c} give
good results even between the division points $z_i$ (see
Sec.~\ref{sec:cnt-pmf-SMD}). This means that $N$ does not need to be a large
number, and therefore the computational overhead due to the intermediate
equilibrations can be significantly reduced.

\section{PMF of water  molecules in SWNT}
\label{sec:res}
In this section we calculate the PMF that guides the translocation of water
molecules across a periodic structure of densely-packed SWNTs, as well as, the
corresponding position dependent diffusion coefficient.
%
\begin{figure}[Ht]
  \centering
\includegraphics[width=3.2in]{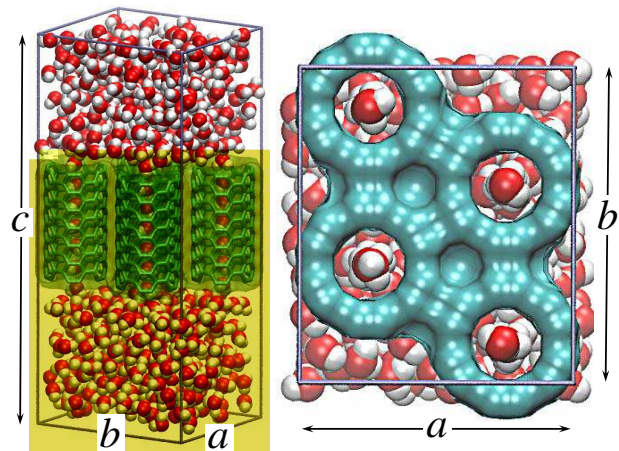}
\caption{(Color online) Lateral (left) and top (right) view of the unit cell
  ($a=20$~\AA, $b=23$~\AA~ and $c=52.5$~\AA) of the simulated water and SWNT system
  using Van der Waals and surface representations. Water molecules cross the SWNTs
  in single files. Figure rendered with VMD\cite{HUMP96}.}
  \label{fig:cnt}
\end{figure}
The choice of this system as a testing ground for our FR method was motivated by
the following. First, water filled SWNTs are nontrivial many particle systems
comprising thousands of atoms, yet they are easy to simulate and the PMF of waters
inside the SWNTs can be easily tuned by changing the Van der Waals interaction
parameters between the carbon and water molecules\cite{hummer01-188}. Second,
SWNTs are hydrophobic nanopores that can be regarded as simplified models for the
much more complex channel proteins. Thus, they are ideal for testing new
computational methods and hypothesis that later can be applied to protein
channels. Finally, during the past few years, SWNT have been intensively studied
through MD simulations revealing many interesting and surprising properties%
\cite{hummer01-188,berezhkovskii02-064503,waghe02-10789,kalra03-10175,%
zhu04-224501,zhu03-236,lu05-11461,dzubiella04-5001}.
In particular, these simulations revealed that hexagonally packed (6,6) SWNTs,
with diameter of 8.1~\AA, spontaneously fill with a single file of water molecules
when connecting two water reservoirs. Water molecules diffuse across the tubes in
a concerted fashion, with a diffusion rate close to the corresponding bulk value.
This correlated motion can be described rather well with a continuous-time random
walk (CTRW) model\cite{berezhkovskii02-064503}. As an alternative to the CTRW
model, here we propose a more general stochastic model in which the motion of each
water molecule along the $z-$axis of a SWNT is characterized by an effective
(position dependent) diffusion coefficient $D(z)$ and a PMF, $U(z)$. Both
quantities can be determined efficiently and simultaneously by our FR method.

We consider a periodic system (see Fig.~\ref{fig:cnt}) of 4 hexagonally-packed
identical SWNTs of (6,6) armchair type. Each SWNT (156 atoms) has a C$-$C diameter
of $8.2~\text{\AA}$ and length $14.7~\text{\AA}$.  On both sides of the SWNTs
there is a water layer of width $18.9~\text{\AA}$. The system contains $556$ water
molecules in total.  The unit cell has dimensions $23 \times20 \times
52.5~\text{\AA}^3$ and contains a total of $2292$ atoms.
All MD simulations were performed in the NpT ensemble ($T=300$~K and $p=1$~atm),
using periodic boundary conditions and the PME method for full
electrostatics\cite{essmann95-8577}. Water molecules were modeled as
TIP3P\cite{jorgensen83-926}.
To facilitate the comparison between the PMFs obtained with different methods, the
Van der Waals parameters of the C atoms (of type CA for benzene in the CHARMM
force field)\cite{MACK98} were changed (from $\epsilon = 0.10$ to $\epsilon =
0.13~\text{kcal/mol}$, and from $R_0=3.76$ to $R_0=4.81~\text{\AA}$, respectively)
to artificially increase the size of the potential barriers in the PMF from $0.35$
to $2~\text{k$_BT$}$.
All simulations were performed with the program NAMD2\cite{KALE99}, with a
performance of $\sim$~1~day/ns on 8 CPUs of a G4 Beowulf cluster (preferred for
repeated SMD pullings), or $\sim$~12~hours/ns on 24 CPUs (preferred for long EMD
simulations).
Just like in previously reported
simulations\cite{zhu03-236,kalra03-10175,berezhkovskii02-064503,hummer01-188}, the
initially empty SWNTs filled up completely with water (i.e., 5 molecules per
nanotube) in the first few hundreds of ns. Also, the arrangement of the SWNTs
prevented water molecules from entering the space between them.

\subsection{PMF from equilibrium MD simulations}
\label{sec:cnt-pmf-EMD}
\begin{figure}[Ht]
  \centering
\includegraphics[width=3.2in]{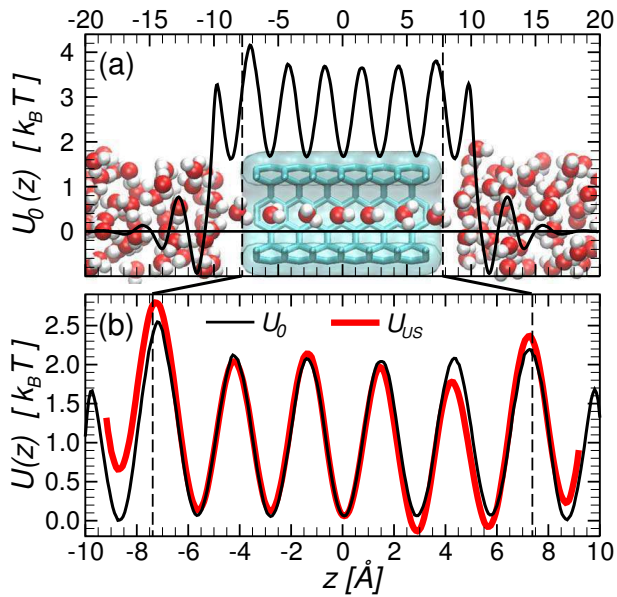}
\caption{(Color online) (a) PMF $U_0(z)$ of a water molecule along the $z$-axis of
  one of the SWNTs obtained through equilibrium MD simulations. The included
  snapshot illustrates a completely filled SWNT with five water molecules located
  about the corresponding PMF minima. (b) Comparison between $U_0(z)$ (thin curve)
  and the same PMF $U_{US}(z)$ (thick curve) obtained from umbrella
  sampling. Graphics rendered with the program VMD\cite{HUMP96}.}
  \label{fig:cnt-pmf}
\end{figure}

The PMF $U_0(z)$ [Eq.~\eqref{eq:pmf0}] was determined from a 9~ns long EMD
trajectory recorded after the system was equilibrated. The histogram $p_0(z)$ was
constructed by binning the $z$-coordinate of the O-atoms of all water molecules.
No visible change in the normalized distribution $p_0(z)$ could be noticed when
the first 7~ns part of the EMD trajectory was used to build it, indicating
that the sampling was complete. Inside the SWNTs (see Fig.~\ref{fig:cnt-pmf}a)
$U_0(z)$ has five equidistant minima (water binding sites) with separation
distance $2.8~\text{\AA}$ and almost identical potential barriers of height
$2~k_BT$. It is convenient to label these minima from 1 to 5 along the positive
$z$-direction.
On both sides, moving away from the SWNTs into the bulk water the PMF exhibits
three more minima (labeled $0,-1,-2$ and $6,7,8$, respectively) before it flattens
out. Water molecules to move in an out the SWNTs [i.e., to hop between minima
$(0,1)$ and $(5,6)$] must overcome roughly the same energy barrier as the ones
located inside the tubes. However, there is a strong spatial inhomogeneity of the
water distribution right outside the nanotubes that is related to the large
asymmetry of the energy barrier connecting minima $(-1,0)$ and $(6,7)$,
respectively. The PMF profile is reflected by the snapshot of the water molecules
in Fig.~\ref{fig:cnt-pmf}a and is compatible with the observation that single-file
water transport through SWNTs usually occurs in unidirectional bursts.
We have also determined the PMF, $U_{US}(z)$, inside the SWNTs by using umbrella
sampling and WHAM, as described in Sec.~\ref{sec:us-wham}. A total of six sampling
windows were used. For convenience, these were centered, by means of HGPs with
$k=1.2~\text{kcal/mol}\cdot\text{\AA}^2$, on the six maxima within the SWNTs of $U_0(z)$. The
samplings of the biased systems were carried out through 5~ns long EMD
simulations. To speed up the computation, the HGPs in the four SWNTs were centered
on different maxima. Thus each EMD trajectory provided four biased distribution
histograms $p_i(z)$. The fact that these were properly sampled was tested by
making sure that the histograms corresponding to the first 4~ns part of the
EMD trajectory coincided with the one obtained from the entire
trajectory. Finally, $U_{US}(z)$ was determined by solving the WHAM
Eqs.~\eqref{eq:wham}. As shown in Fig.~\ref{fig:cnt-pmf}b, the agreement between
the calculated $U_0(z)$ and $U_{US}(z)$ is rather good, though not perfect. 

\subsection{PMF from nonequilibrium cv-SMD pullings}
\label{sec:cnt-pmf-SMD}

Next, by employing our new FR approach described in Sec.~\ref{sec:f-r-smd}, the
PMF $U_{FR}(z)$ was determined from a small number of fast F and R cv-SMD pullings
of water molecules across the SWNTs.
In each cv-SMD simulation four water molecules were pulled across the SWNTs (one
molecule per nanotube) by applying a stiff ($k=10~\text{kcal/mol} \cdot
\text{\AA}^2$) HGP [see Eq.~\eqref{eq:hgp}] that moved with $v=20$~\AA/ns along
the $z$-axis of the nanotubes. Only four such pullings were performed in both F
and R directions between the extremities of the interval $z\in [-10,10] $~\AA.
Each cv-SMD simulation was started from an equilibrated configuration (in
accordance with the applicability of Crooks' TFT) and was 1~ns long. Out of the
$4\times 4 = 16$ F and R trajectories only those where retained for analysis in
which the corresponding SWNT remained filled with water at all times. In several
cases, once the pulled water molecule crossed halfway the channel the binding
sites behind it remained unoccupied. Since such configurations correspond to a
different free energy profile, such trajectories must be dropped in determining
the PMF for a completely filled SWNT. Thus, we ended up with 7 F and 14 R paths
for calculating the PMF.
\begin{figure}[Ht]
  \centering
\includegraphics[width=3.2in]{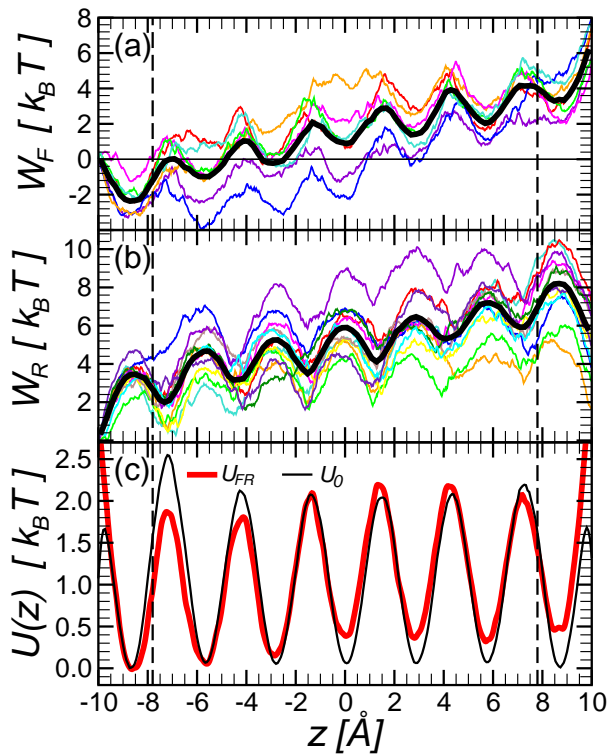}
\caption{(Color online) Work along (a) forward and (b) reverse SMD pullings. The
  mean work $\overline{W}_{F/R}$ is shown as a thick solid curve. (c) Comparison
  between $U_0(z)$ (thin curve) and $U_{FR}(z)=\left(
    \overline{W}_F-\overline{W}_R \right)/2$ (thick curve), obtained from fast
  forward and reverse SMD pullings. Vertical dashed lines indicate the extremities
  of SWNTs.}
  \label{fig:pmf-cnt}
\end{figure}
Because we already know the ``exact'' PMF $U_0(z)$, we deliberately did not choose
to add more trajectories from extra simulations.  Indeed, since in the case of
large biomolecules one can afford only a small number of SMD runs, our goal here
is to test the viability of the proposed FR method for calculating PMFs under such
unfavorable conditions.
The external work along the F and R paths, including the mean work
$\overline{W}_{F/R}$, are shown in Fig.~\ref{fig:pmf-cnt}a and b, respectively.
Note that in order to display $W_R$ on the same plot with $W_F$, the sign of the
former needs to be reversed and shifted to the origin of the latter.
As shown in Fig.~\ref{fig:pmf-cnt}c, within the SWNTs (indicated by dashed
vertical lines) the PMF $U_{FR}=(\overline{W}_F - \overline{W}_R)/2$ agrees
surprisingly well with $U_0$ and $U_{US}$.
We have checked (results not shown) that by increasing the pulling speed to
$v=40~\text{\AA}$ and using a similarly small number of F and R trajectories, the
quality of the obtained PMF is very similar to the one shown in
Fig.~\ref{fig:pmf-cnt}c.  However, in this particular case, the higher the pulling
speed the most likely that the SWNTs will partially empty during pulling and,
therefore, more runs are necessary to collect a minimum number of paths for
calculating the PMF.

\begin{figure}[Ht]
  \centering
\includegraphics[width=3.2in]{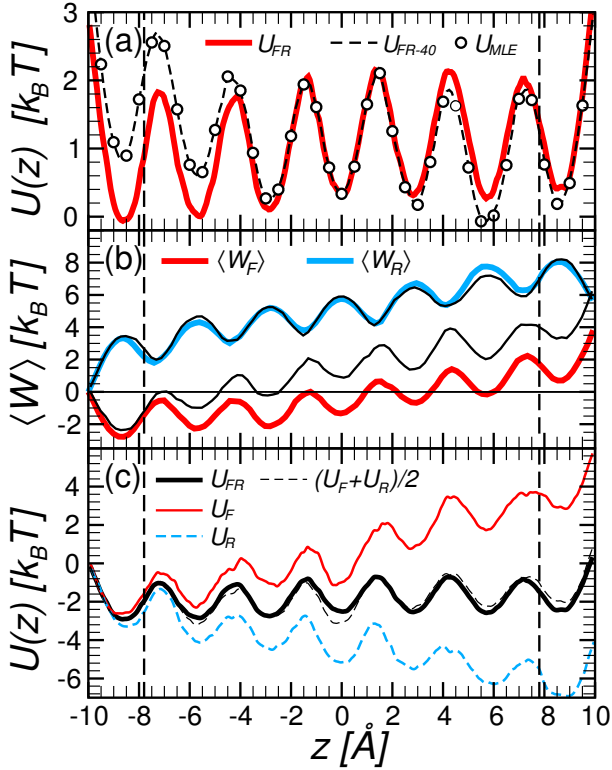}
\caption{(Color online) (a) PMF of a water molecule in a SWNT determined by using
  the FR method: channel as a whole (solid) and divided in 40 adjacent
  segments of the same length (dashed). The PMF at the ends of the segments
  obtained with the MLE method are also shown (circles). (b) Mean forward and
  reverse work for SWNTs considered as a single segment (thick-solid) and as 40
  adjacent segments (thin-solid).  
  (c) PMFs calculated within the cumulant approximation considering only forward
  (thin-solid) and reverse (thick-dashed) pullings. The arithmetic mean of these
  two (thin-dashed) matches almost perfectly the PMF from the FR method
  (thick-solid). 
  Vertical dashed lines indicate the extremities of SWNTs.}
  \label{fig:pmf-smd}
\end{figure}

As discussed in Sec.~\ref{sec:smd-tft-je}, for TFT to be valid it is necessary
that the initial states of both F and R pullings be sampled from an equilibrium
distribution. Thus, strictly speaking the above results using the FR method
applies only to the two ends of the considered interval. The good agreement
between $U_{FR}$ and $U_0$ suggests that our method may give reliable PMFs
for all values of the RC $z$ in the considered interval.
However, as shown next, it is simple to extend our FR method to cases where this
issue may impact negatively on the determination of the PMF.
Thus, the RC interval was divided into 40 segments of the same length. For each
division point, the system was equilibrate for a few hundreds of ns by using the
same HGP centered about those points. Starting from statistically independent
equilibrium configurations, 4 pullings with the same $v=10$~\AA/ns in both F and R
directions were carried out on each segment. None of the SWNT emptied during these
short cv-SMD runs and, therefore, all trajectories were used for analysis. The
resulting PMF, $U_{FR-40}(z)$, is shown in Fig.~\ref{fig:pmf-smd}a.  The agreement
with the previously determined $U_{FR}$ is fairly good, especially inside the
SWNTs.  Closer inspection suggests that compared to the ``exact'' $U_0$,
$U_{FR-40}$ is not as good as $U_{FR}$. Thus, one may conclude, that more sampling
in the FR method does not necessarily give better results.  Indeed, in the FR
method we only need a good estimate of the mean F and R work, and not a complete
sampling of the corresponding work distribution functions.  However, it is very
difficult to estimate how good is the mean work calculated from a few fast
pullings.
Also, we have calculated the PMF in the division points by using the MLE method,
as shown in Fig.~\ref{fig:pmf-smd}a. The $U_{MLE}$ points fall right on the
$U_{FR-40}$ curve, suggesting again that in the FR method the quality of the
sampled paths is more important than the optimal statistical analysis of the
trajectories.

In Fig.~\ref{fig:pmf-smd}b the mean F and R work is plotted for both cases when
the F/R pullings are done in one shot (thick curves) and on the segments
separately (thin curves). While $\overline{W}_R$ for both cases match almost
perfectly, the difference between the corresponding mean F work is quite
significant and most definitively is the source of discrepancy between $U_{FR}$
and $U_{FR-40}$. This difference may be due to the smaller number of F
trajectories used in case one, or to partially emptied sites towards the ends of
the SWNTs during the simulations along the segments that were not accounted for
properly. However, it is worth noticing that the mismatch between the PMFs is less
pronounced than for the mean F work.

In any event, for the same SMD data, the FR method gives far better results than
the currently used cumulant approximation method based on JE (see
Sec.~\ref{sec:pmf-smd-je}). In Fig.~\ref{fig:pmf-smd}c the PMFs determined by
applying the cumulant approximation separately to F and R trajectories, i.e.,
$U_F$ and $U_R$, are compared to $U_{FR}$. It is clear that both $U_F$ and $U_R$
are biased in opposite directions. Apparently this behavior was recognized in
previous work in which the PMF of a glycerol molecule in a GlpF channel was
calculated for the first time. To eliminate the bias from only F pullings, the
authors partitioned the GlpF channel into 12 segments and artificially applied in
an alternating fashion F and R pullings in adjacent segments. Our FR approach for
determining PMFs naturally solves this biasing issue due to the invalidity of JE
for few, fast unidirectional SMD trajectories.
We also note that the arithmetic mean of $U_F$ and $U_F$ (Fig.~\ref{fig:pmf-smd}c)
matches rather well $U_{FR}$ indicating that in fact the 2nd cumulant correction
of the work to the PMF is irrelevant in the FR method, in which the mean
dissipative work $\overline{W}_d\gg\sigma_W^2/2$ is already correctly accounted
for by combining F with R paths.


\subsection{Dissipative work and diffusion coefficient}
\label{sec:Wd-D}

Next, we focus on the determination of the mean dissipative work and the
corresponding diffusion coefficient. 
\begin{figure}[Ht]
  \centering
\includegraphics[width=3.4in]{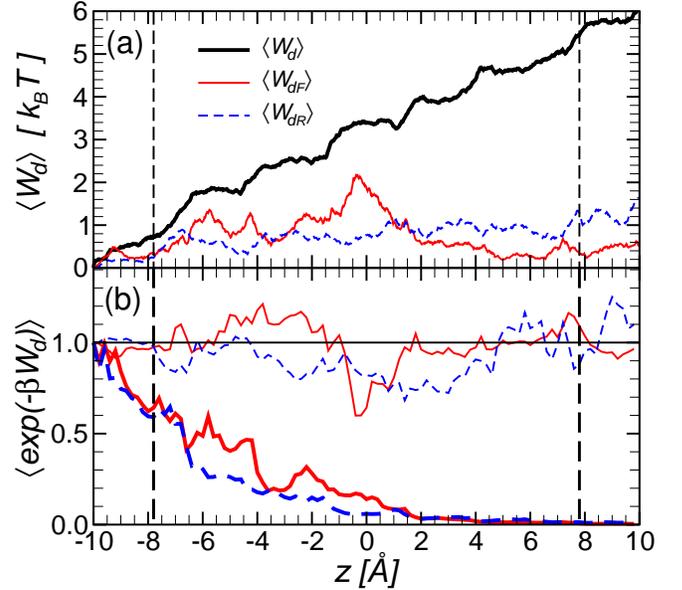}
\caption{(Color online) (a) Mean dissipative work 
  determined from the FT method (thick-solid) and the cumulant approximation
  applied separately to forward (thin-solid) and reverse (thin-dashed) pullings.
  (b) Validity test of JE along forward (solid) and reverse (dashed) processes.
  The PMF $U$ in the corresponding dissipative work $W_{d\,F/R}=W_{F/R}\mp U$ is
  determined from the FR (thick) and the cumulant approximation (thin) methods,
  respectively.}
  \label{fig:wd}
\end{figure}
In Fig.~\ref{fig:wd}a the mean dissipative work derived from the individual F/R
pullings and from the FR method are plotted. As expected, $\overline{W}_{d\,F/R} =
\sigma_{F/R}^{2}/2$ calculated from the variance of $W_{F/R}$ seriously
underestimate $\overline{W}_d$ determined from the FR method by using
Eq.~\eqref{eq:fr-2c}.
This observation has several consequences.
First, the fact that $\overline{W}_{d\,F/R}$ does not increase fast enough with
the pulling distance clearly indicates that only a small region about
$\overline{W}_{F/R}$ of $P_{F/R}(W)$ is sampled and not the entire work
distribution function.
Second, the strongly biased PMFs $U_{F/R}$, obtained from the cumulant
approximation, lead to underestimated dissipative work $W_{d\,F/R}=W_{F/R}\mp
U_{F/R}$ that give the false impression that the JE equation is satisfied along
the F/R pullings, as shown in Fig.~\ref{fig:wd}b (thin curves). This, of course,
is expected because $U_{F/R}$ are calculated based on the assumption that JE
holds.  The reality is that, in fact, JE fails to hold for both F and R pullings
as the system departs from equilibrium. The reason, of course, is that paths with
negative dissipative work ($W_d<0$) that are crucial for the validity of JE
(Eq.~\eqref{eq:JEs}) are not sampled. This is clearly illustrated in
Fig.~\ref{fig:wd}b where $\left\langle \exp(-W_{d\,F/R}) \right\rangle$, plotted
by using the correct expressions $\overline{W}_{d\,F/R} = \overline{W}_{F/R} \mp
\Delta{U}$ (thick curves), decay rapidly towards zero as the system is pulled away
from equilibrium. Clearly, the larger the deviation from equilibrium the less JE
is satisfied.

The position dependent $D(z)$ can be calculated from the slope of $\overline{W}_d$
according to Eq.~\eqref{eq:fr-diff}. Since the mean dissipative work is almost
linear it is not surprising that the diffusion coefficient has an almost constant
value $D\approx 71$~\AA$^2$/ns. This is more than three times smaller than the
bulk diffusion coefficient of water $D_{\text{bulk}}\approx 250$~\AA$^2$/ns.

\subsection{Stochastic model of water transport in SWNTs}
\label{sec:FPE}

The determined $U(z)\equiv U_{FR}(z)$ and $D$ provide the input in the FPE
Eq.~\eqref{eq:fr-FP} for describing water transport through SWNT on meso/macro
time scales. This should be regarded as a generalization of CTRW model of
Berezhkovskii and Hummer \cite{berezhkovskii02-064503}.  In principle, by solving
the FPE for the nonequilibrium distribution function $p(z,t)$ for well defined
initial and boundary conditions one can completely characterize the single-file
transport of water molecules in the considered SWNTs.  A detailed analysis along
this line will be reported in another publication.

In the CTRW model single-file water molecules occupy the binding sites (PMF
minima) within the SWNT. Since they cannot pass each other, the diffusion of water
molecules across the nanotube is brought about by random hops to the empty binding
sites right in front or behind them.  The waiting (or residence) time between two
consecutive hops is a stochastic Poisson process. Besides the equidistant spacing
between two adjacent sites $a$, the mean waiting time $\tau$ is the defining
parameter of the CTRW model. In terms of $\tau$ the effective diffusion
coefficient is $D_{\text{eff}}=a^2/2\tau$.

In our stochastic model $\tau$ is identified with the mean first passage time\cite{risken96}
(MFPT) from one minimum ($z_i$, $i=1,\dots,5$) of the PMF $U(z)$ into the adjacent
one $z_j$, with $j=i\pm 1$, and is given by
  \begin{equation}
    \label{eq:mfpt}
    \tau_{i,j} = \int_{z_i}^{z_j} dx\,e^{U(x)}/D(x) \int_{z_i}^x dy\, e^{-U(y)} \;.
  \end{equation}
Now, the mean waiting time can be expressed as
\begin{equation}
  \label{eq:mwt}
  \overline{\tau} = \left(\sum_{i=1}^{N-1} \tau_{i,i+1} + \sum_{i=2}^N
    \tau_{i,i-1}\right)/2(N-1)\;.
\end{equation}
In our case $N=5$ and the corresponding mean waiting time $\overline{\tau}\approx
84~\text{ps}$. Applying our stochastic model to the pristine SWNT considered in
Ref.~\onlinecite{berezhkovskii02-064503} (for which the barrier height between
binding sites is only $0.35~k_BT$ compared to $2~k_BT$ in our modified SWNTs) one
obtains $\overline{\tau}\approx 12.9$~ns that compares very well with the reported
$13$~ns.

Furthermore, the effective diffusion coefficient $D_{\text{eff}}$ of single-file
water molecules in SWNTs can be defined as
\begin{equation}
  \label{eq:Deff}
  D_{\text{eff}}=D \left(\overline{a}^2/2D \overline{\tau}\right) \;,
\end{equation}
where $\overline{a}=2.8~\text{\AA}$ is the mean spacing between two adjacent
binding sites. $D_{\text{eff}}$ describes the diffusion of fictitious particles in
the absence of the PMF with the same mean diffusion time on a distance
$\overline{a}$ as the mean waiting time $\overline{\tau}$. In our case we get
$D_{\text{eff}}\approx 45~\text{\AA}^2/ns$. It is this diffusion coefficient that
can be measured from the well known asymptotic formula $\left\langle \Delta z^2(t)
\right\rangle = 2D_{eff}t$ from EMD simulations. Indeed, from our simulations we
obtain $D_{\text{eff}}\approx 48~\text{\AA}^2/ns$, in very good agreement with the
result from our stochastic model.

Finally, one can calculate the mean permeation time $T$ across the channel in two
different ways: (i) as the MFPT from one end of the nanotube to the other, and
(ii) as $L^2/2D_{eff}$, where $L$ is the length of the SWNT. In both cases one
obtains essentially the same result: $T\approx 1.45$~ns between $z_1$ and $z_5$,
and $T'\approx 3.2$~ns between $z_0$ and $z_6$ (i.e., between the binding sites
right outside the ends of the SWNTs).
The observed 12 permeations per nanotube in 9~ns corresponds to a permeation time
1.38~ns that is a good estimate for $T$ but it is considerably shorter than $T'$.
Thus, even in this relatively simple case very long EMD simulations are needed to
calculate the unidirectional water flux through the modified SWNTs by simply
counting the number of full permeations of water molecules, reinforcing once again
the value of our stochastic modeling approach.

\section{Conclusions}
\label{sec:conc}

The potential and value of Crooks' TFT for determining free energy profiles is
becoming more apparent both theoretically\cite{shirts03-140601} and
experimentally\cite{collin05-231}. 
In this paper we have shown that by employing Crooks' TFT\cite{crooks00-2361}
within the stiff spring approximation the potential of mean force along a suitably
chosen reaction coordinate can be determined (at least semiquantitatively) from
combining a few fast forward and time reversed nonequilibrium processes started
from an equilibrium configuration and subject to the same evolution protocol of
the reaction coordinate.
In the proposed FR method one determines simultaneously both the PMF ($U$) and the
mean dissipative work ($\overline{W}_d$) without invoking JE. In fact, JE is not
even satisfied for fast F or R pullings simply because processes with negative
dissipative work (that transiently violates the second law and are exponentially
small in number) are not sampled.
The FR method is based on a key observation involving Crooks' TFT (which is more
general than JE): whenever the F work distribution function $P_F(W)$ is Gaussian
(e.g., in the case of the stiff-spring approximation) then $P_R(W)$ is also
Gaussian. Furthermore, $P_{F/R}(W)$ have the same width and are shifted by
precisely twice the corresponding free energy difference between the equilibrium
states connected by the F and R processes. Thus, both $U$ and $\overline{W}_d$ can
be readily determined from the mean F and R work ($\overline{W}_{F/R}$). The
practical success of the FR method stems from the fact that the mean work
$W_{F/R}$ can be measured rather accurately from only a few fast F/R pullings.
This also explains why previous methods, based on the direct application of JE,
fail to work away from equilibrium, making them inefficient for practical
applications. Indeed, the width of $P_{F/R}(W)$, which is proportional to
$\overline{W}_d$, cannot be determined even approximately from a few
unidirectional pullings, unless these are close to equilibrium and rendering
$P_{F/R}(W)$ sufficiently narrow.
This FR method works rather well for both small and large (e.g., biomolecular)
systems. Although here we applied and tested the FR method in the context of SMD
simulations, in principle this can be applied equally well to analyze properly
designed single molecule experiments. 

To test its viability, we have applied the FR method to determine the PMF and
position dependent diffusion coefficient of single-file water molecules in SWNTs.
The derived PMF was found to be in good agreement with the one obtained from
standard EMD methods, e.g., umbrella sampling. In case of large biomolecular
systems, when EMD methods become computationally unaffordable, the proposed FR
method may provide the only hope for determining PMFs. In addition, the FR method
has the unique feature that it determines simultaneously both the PMF and the
corresponding position dependent diffusion coefficient. These two quantities then
can be used in a stochastic model that permits the study of the dynamics of the
system along the reaction coordinate on meso/macro time scale by retaining its
microscopic spatial resolution. For example, our stochastic model provides a
generalization of the recently proposed CTRW model for single-file water transport
in SWNTs\cite{berezhkovskii02-064503}.

\section{Acknowledgments}
This work was supported in part by grants from the University of
Missouri Research Board, the Institute for Theoretical Sciences, a joint institute
of Notre Dame University and Argonne National Laboratory, the U.S. Department
of Energy, Office of Science through contract No.~W-31-109-ENG-38, and NSF through
FIBR-0526854. 


\end{document}